\documentclass[cits]{PoS}
\usepackage{epsfig,macros}
\usepackage{amssymb,amsmath}
\usepackage{booktabs,dcolumn}
\newcolumntype{C}{>{$}c<{$}} 
\newcommand{\onecol}[2]{%
\begin{minipage}[t]{#1}{#2\vfill}\end{minipage}}
\title{%
Towards a non-perturbative matching of HQET\\ 
and QCD with dynamical light quarks%
\thanks{Based on the contributions of J.\,Heitger and P.\,Fritzsch.}
}

\author{%
\epsfxsize=2.5 true cm
\epsfbox{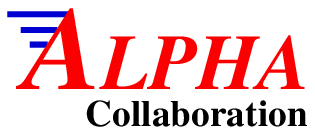}\vspace{-2.5cm}\hfill
{\onecol{3.15cm}{\vspace{-0.925cm}\fns\it%
CERN-PH-TH/2007-171\\MS-TP-07-32\\MIT-CTP 3878\\DESY 07-166\\SFB/CPP-07-61}}
}

\ShortTitle{%
Towards a non-perturbative matching of HQET and QCD 
with dynamical light quarks
}

\author{%
Michele Della Morte\\
CERN, Physics Department, TH Unit,
CH-1211 Geneva~23, Switzerland\\
E-mail: \email{dellamor@mail.cern.ch}
}
\author{%
Patrick Fritzsch, Jochen Heitger\\
Westfälische Wilhelms-Universität Münster, 
Institut für Theoretische Physik,\\
Wilhelm-Klemm-Strasse~9, D-48149 Münster, Germany\\
E-mail: \email{fritzsch@uni-muenster.de,heitger@uni-muenster.de}
}
\author{%
Harvey B. Meyer\\
Massachusets Institute of Technology, Center for Theoretical Physics,\\
Cambridge, MA 02139, U.S.A.\\
E-mail: \email{meyerh@mit.edu}
}
\author{%
Hubert Simma\thanks{
Present address: 
Universit\`a di Milano ``Bicocca'', Dip. di Fisica,
Piazza della Scienza 3, I-20126 Milano, Italy
}\;, Rainer Sommer\\
Deutsches Elektronen-Synchrotron DESY, Zeuthen,\\
Platanenallee~6, D-15738 Zeuthen, Germany\\
E-mail: 
\email{hubert.simma@desy.de,rainer.sommer@desy.de}
}

\abstract{%
We explain how the strategy of solving renormalization problems in HQET 
non-perturbatively by a matching to QCD in finite volume can be implemented 
to include dynamical fermions. 
As a primary application, some elements of an HQET computation of the mass 
of the b-quark beyond the leading order with $N_{\rm f}=2$ are outlined. 
In particular, the matching of HQET and QCD requires relativistic QCD 
simulations in a volume with $L \approx 0.5\,{\rm fm}$, which will serve to 
quantitatively determine the heavy quark mass dependence of heavy-light 
meson observables in the continuum limit of finite-volume two-flavour 
lattice QCD.
As a preparation for the latter, we report on our determination of the 
renormalization constants and improvement coefficients relating the 
renormalized current and subtracted bare quark mass in the relevant weak 
coupling region. 
The calculation of these coefficients employs a constant physics condition 
in the Schr\"odinger functional scheme, where the box size $L$ is fixed by 
working at a prescribed value of the renormalized coupling. 
}

\FullConference{%
The XXV International Symposium on Lattice Field Theory\\
July 30 - August 4 2007\\
Regensburg, Germany
}
\begin{document}
\section{Introduction}
\label{Sec_intro}
In the light of the expected progress in flavour physics thanks to the
impending B-physics experiments \cite{Barsuk:2005ac,Akeroyd:2004mj},
precision lattice QCD more and more becomes to play a crucial r\^{o}le for 
a quantitative and accurate interpretation of these experimental results 
in the framework of the Standard Model and beyond, since it provides a 
theoretically sound approach to non-perturbatively compute the 
contributing matrix elements of operators among hadronic states.

A particular problem of dealing with heavy-light systems involving the 
b-quark as the heavy flavour by means of lattice QCD consists in the
two disparate intrinsic scales that actually accompany any lattice 
calculation: 
the lattice spacing, $a$, has to be much smaller than $1/m_{\rm b}$ in order
to allow for a fine enough resolution of the B-meson states in question,
and the linear extent of the lattice volume, $L$, has to be large enough for 
finite-size effects to be under control.
Heavy Quark Effective Theory (HQET) on the 
lattice~\cite{stat:eichten,stat:eichhill1}, however, which relies upon a
systematic expansion of the QCD action and correlation functions in inverse 
powers of the heavy quark mass ($m$) around the static limit, offers a 
formally reliable solution to this problem.
Still, for lattice HQET and its numerical applications to lead to precise 
results with controlled systematic errors in practice, two shortcomings had 
to be left behind first.

One is the exponential growth of the noise-to-signal ratio in static-light 
correlation functions, which is a consequence of the appearance of power 
divergences in the effective theory. 
As demonstrated in studies in the quenched approximation
\cite{fbstat:pap1,HQET:statprec,ospin:Nf0,bb:pap1,fbstat:pap2} as well as in 
the theory with $N_{\rm f}=2$ dynamical quarks~\cite{zastat:Nf2}, this 
problem can be overcome by a clever modification of the traditional 
Eichten-Hill discretization of the static action.

Another difficulty, more serious on the theoretical level, is associated 
with the aforementioned power divergences.
Since in the effective theory mixings among operators of different 
dimensions are present, already the static limit of HQET is affected by a 
power-law divergent ($\sim g_0^2/a$) additive mass renormalization.
Unless the theory is renormalized non-perturbatively \cite{Maiani:1992az},
it follows from this power-law divergence of lowest-order HQET 
--- and, of course, from further ones $\sim g_0^2/a^{n+1}$ that arise at 
$\Or(1/m^n)$, $n\ge 1$ --- that the continuum limit does not exist owing to 
a remainder, which, at any finite 
order \cite{stat:eichhill_1m,mbstat:dm_MaSa,mbstat:dm_DirScor} 
in perturbation theory, diverges in the continuum limit.

In~\Ref{HQET:pap1} a general solution to the latter has been worked out and 
numerically implemented for a determination of the b-quark's mass in the 
static and quenched approximations as a test case.
The method is based on a \emph{non-perturbative matching of HQET and QCD in 
finite volume}.
It was subsequently extended to also include the $\rmO(\minv)$ terms into 
the quenched computations of the mass of the b-quark, 
$\mbbMS(\mbbar)$~\cite{HQET:mb1m} 
(see~\Refs{reviews:NPRrainer_nara,lat07:michele} for recent reviews in
broader context), and of the B-meson decay constant~\cite{lat07:nicolas}.

An attractive property of the strategy, briefly summarized
in~\Sect{Sec_survey}, is that most parts of the actual calculation do not 
involve very large lattices.
Hence, it is natural to remove the quenched approximation as the dominating 
remaining systematic uncertainty in our previous works using this method.
The additional computational effort required if dynamical quarks are 
included is only moderate, except for the last step that involves the
extraction of B-meson properties from simulations in physically large 
volumes (with spatial extents of $\approx 2\,\Fm$ or more) and thus will be 
computationally much more demanding than the finite-volume simulations for
the non-perturbative renormalization part.

In the present report we outline the various steps towards an HQET 
computation of the mass of the b-quark including the 
$\rmO(\minv)$ correction along the lines 
of~\Refs{HQET:pap1,HQET:mb1m} in two-flavour lattice QCD, where most of the
emphasis is put on the renormalization of the effective theory through the
non-perturbative matching to QCD in finite volume in order to perform the
power-divergent subtractions.
This step requires, in particular, a determination of the relation between 
the renormalization group invariant (RGI) and the subtracted bare heavy
quark mass in the relevant parameter region of $N_{\rm f}=2$ QCD, which we 
present together with numerical results on the corresponding renormalization
constant and improvement coefficients in some detail in~\Sect{Sec_bXz}.
Results from the matching itself, which has just been started at the time of 
writing, as well as from the necessary simulations of the effective theory 
in small and intermediate volumes will only be available at later stages of 
our project.
\section{Survey of the computational steps}
\label{Sec_survey}
%
%
\begin{figure}[htb]
\centering
\leavevmode
\epsfig{file=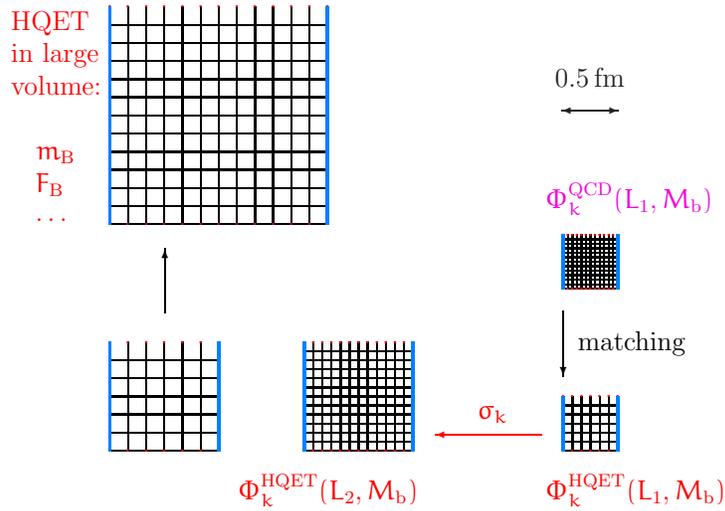,width=0.625\textwidth}
\caption{%
The strategy for performing computations in lattice HQET via a
non-perturbative determination of the HQET parameters from QCD simulations 
in a small volume. 
It is designed such that steps indicated by arrows are to be repeated at 
smaller lattice spacings to reach a continuum limit.
}\label{fig:strat}
\end{figure}
Let us briefly recall the general strategy, introduced in \cite{HQET:pap1}.
It allows for a formulation of (zero-velocity) HQET in the framework of 
lattice QCD, where all steps of the computations including the 
renormalization are carried out non-perturbatively and the continuum limit 
can be taken.

The basic idea is illustrated in~\fig{fig:strat} and starts from a finite 
volume of extent $L_1\approx 0.5\,\Fm$.
There, one chooses lattice spacings $a$ sufficiently smaller than 
$1/\mbeauty$ such that the b-quark propagates correctly up to controllable 
discretization errors of order $a^2$. 
Since the relation between the RGI mass and the bare mass in QCD is
known \cite{msbar:Nf2}, suitable finite-volume observables $\Phi_k(L_1,\Mb)$ 
can be calculated as a function of the RGI b-quark mass, $\Mb$, and 
extrapolated to the continuum limit. 
The next step is to perform the power-divergent subtractions 
non-perturbatively by a set of matching conditions, in which the results
obtained for $\Phi_k$ are equated to their representation in HQET
(r.h.s.~of~\fig{fig:strat}).
At the same physical value of $L_1$ but for resolutions $L_1/a=\rmO(10)$, 
the previously computed heavy-quark mass dependence of $\Phi_k(L_1,\Mb)$ in 
finite-volume QCD may be exploited to determine the bare parameters of the 
effective theory for $a\approx(0.025-0.05)\,\Fm$.
In order to evolve the HQET observables to large volumes, where contact with 
experiments can be made, one also computes them at these lattice spacings in 
a larger volume, $L_2=2L_1$.
The resulting relation between $\Phi_k(L_1)$ and $\Phi_k(L_2)$ is encoded in
associated step scaling functions $\sigma_k$, as indicated 
in~\fig{fig:strat} as well.
Finally, the knowledge of $\Phi_k(L_2,\Mb)$ and employing resolutions 
$L_2/a=\rmO(10)$ fixes the bare parameters of the effective theory
for $a\approx(0.05-0.1)\,\Fm$ so that a connection to those lattice spacings 
is established, where large-volume observables, such as the B-meson mass or 
decay constant, can be calculated (l.h.s.~of~\fig{fig:strat}). 

Having in mind the computation of $\Mb$ as the specific application,
this sequence of steps yields an expression of $\mB$ (taken to be the 
physical input) as a function of $\Mb$ via the quark mass dependence of 
$\Phi_k(L_1,\Mb)$, which eventually can be inverted to arrive at the desired
physical value of the RGI b-quark mass extracted from the effective theory.
As pointed out before, the whole construction is such that its various 
pieces separately have a continuum limit.
With the realization of this strategy for the quenched case it was shown 
in~\Ref{HQET:mb1m} that a determination of $\Mb$ including 
$\rmO(1/m)$ in HQET only requires up to three matching observables, 
$\Phi_1$, $\Phi_2$ and $\Phi_3$, if the spin-averaged B-meson mass is used 
as physical input.
That is also the path we will follow in our present extension to the case of
two-flavour QCD.
\subsection{Definition of the matching volume}
\label{Sec_survey_L1}
We consider QCD with $\nf=2$ mass-degenerate dynamical quarks, which are
identified with up and down.
All other quarks are treated in the quenched approximation.
A particularly convenient renormalization scheme, in which finite-volume
observables suitable for a non-perturbative matching of the effective theory
with QCD can readily be constructed \cite{HQET:pap1,HQET:pap2,HQET:pap3},
is the Schr\"odinger functional (SF) \cite{SF:LNWW}.
Relativistic and static quarks were introduced in \cite{SF:stefan1}
and \cite{zastat:pap1}, respectively, where in the latter reference it was 
found that the HQET expansion of the boundary quark fields is trivial up to 
and including $\minv$--terms.\footnote{
From now on, $m$ generically denotes the mass parameter of the heavy quark 
treated in the effective theory, while the masses of the non-degenerate 
quark flavours in the relativistic theory are distinguished explicitly where 
necessary.
}
Adopting any unexplained notation from \Refs{HQET:pap1,zastat:pap1}, we only
mention the periodicity phase $\theta$ of the fermion fields as a further 
kinematic parameter and the fact that homogeneous Dirichlet boundary 
conditions in time at $x_0=0$ and $x_0=T$ are employed.
Since the parameters $\theta$ and masses of the quenched quarks can be set
independently of those of the sea quarks, the basic situation for 
extracting heavy-light physics from SF correlation functions is the same as
in the quenched approximation \cite{HQET:pap1,HQET:mb1m}.
Moreover, in the finite-volume simulations we set $\theta=0.5$ for the
dynamical light quarks and their PCAC mass to zero, $\ml=0$.

The quantities $\Phi_k$ that enter the non-perturbative matching procedure 
described above have to be evaluated in the continuum limit.
To this end we want to compute them for a series of bare parameters
$(L/a,\beta,\kapl)$ such that the renormalized parameters in the light
quark sector are fixed and thereby physics is kept constant along the 
approach to the continuum limit.
Here, $\kapl$ denotes the hopping parameter of the dynamical light quarks.
Our \emph{constant physics condition} on the renormalized SF coupling, 
$\gbsq(L)$, and the light quark mass reads
\be
\gbsq(L_0)=2.989\,, \quad L_0=\frac{L_1}{2}\,, \quad L_0\,\ml(L_0)=0\,.
\label{cond_L1}
\ee
This choice now \emph{defines} the spatial extent $L_1$ of the volume, in 
which the matching between HQET and QCD is performed.
Although an exact knowledge of $L_1$ in physical units in not yet needed at
this stage, one can already infer from the known running of the SF coupling
for $\nf=2$ \cite{alpha:Nf2_2} that $L_1\approx 0.5\,\Fm$.
Hence, we will finally have $L_2=2L_1\approx 1\,\Fm$ and thus 
$L_{\infty}\equiv 4L_1\approx 2\,\Fm$ for the large volume, which is well 
consistent with the envisaged strategy, \fig{fig:strat}.

We have fixed $\gbsq(L_1/2)=2.989$ by a new simulation at $L_0/a=20$, 
$T=L_0$, and made tentative interpolations in $\beta=6/g_0^2$ for given
$L_0/a\le 16$ to this target value, based on the known dependence of the SF 
coupling and the current quark mass on the bare parameters 
$(\beta,\kappa)$ available from the data of~\Ref{alpha:Nf2_2}.
Using the known $\beta$--function and our experience from the quenched 
calculation \cite{HQET:mb1m}, we can estimate that an uncertainty of about 
$0.04$ in the coupling will translate via the resulting 
one in $L_1$ into an uncertainty in the b-quark mass of at most $0.5\%$.
The condition of zero light quark mass in \eq{cond_L1} is met by setting
$\kappa\equiv\kapl$ to the critical hopping parameter, $\kapc$, estimated 
again on basis of published data \cite{alpha:Nf2_2}, whereby a slight 
mismatch of $|L_0\ml(L_0)|<0.05$ of this condition is tolerable in practice.
The triples $(L_0/a,\beta,\kapl)$, which approximately define the extent 
$L_1$ of the matching volume through \eq{cond_L1} and which are used in our 
subsequent study of improvement and renormalization factors, are collected 
in columns~2~--~4 of \tab{tab:bXzres} in \Sect{Sec_bXz}.  

The preliminary interpolation procedure for $\gbsq(\beta)$ underlying these 
$\beta$--values is currently being checked (and refined) by direct
simulations, in order to avoid a non-negligible systematic error from small 
violations of the condition (\ref{cond_L1}) on the final results.
Yet, this will affect our estimates of $\bm$ and $Z$ in \Sect{Sec_bXz} only 
at a negligible level, because there any deviation from the line of constant 
physics only entails a small change of the $\Or(a^2)$ effects.
\subsection{Fixing the heavy quark mass in finite-volume QCD}
Having fixed $L_1$ via enforcing constant physics at $L_0=L_1/2$, the
computation of the heavy quark mass dependence of the finite-volume 
observables $\Phi_k$, which is the key element in the non-perturbative 
matching step within our strategy, will amount to evaluate heavy-light SF 
correlation functions in a volume $L_1^3\times T$, $T=L_1$, for a series of 
precisely fixed values of the renormalized heavy quark mass covering the 
b-quark mass region.
 
This is achieved by exploiting the $\Or(a)$ improved relation between the 
(subtracted) bare heavy quark mass $\mqh$ and the RGI 
mass~\cite{msbar:Nf2,impr:babp}, viz.
\be
M\,=\,
h(L_0)\,\zm(g_0,L_0/a)\,\,\mqh\left(1+\bm(g_0)\,a\mqh\right)
\,+\,\,\Or\left(a^2\right)\,,
\label{M_mqtil}
\ee 
where
\be
\zm(g_0,L_0/a)=\frac{Z(g_0)\,\za(g_0)}{\zp(g_0,L_0/a)}\,,\quad
a\mqh=\frac{1}{2}\left(\frac{1}{\kaph}-\frac{1}{\kapc}\right)
\label{Zm_mqh}
\ee
and $\za$ is known non-perturbatively from~\Ref{impr:za_nf2}.
The scale dependent renormalization constant $\zp$ may be calculated for 
the relevant couplings on $L_0^3\times T$ lattices with $T=L_0$ in the same 
way as in \Ref{msbar:Nf2}.
The factor
\be
h(L_0)=\frac{M}{\mbar(\mu_0)}=1.521(14)\,,\quad 
\mu_0=\frac{1}{L_0}=\frac{2}{L_1}\,,
\ee
represents the universal, regularization independent ratio of the RGI heavy
quark mass, $M$, to the running quark mass $\mbar$ in the SF scheme at the 
renormalization scale $\mu_0$. 
$h(L_0)$ was evaluated by a reanalysis of the $\nf=2$ non-perturbative quark 
mass renormalization data published in \Ref{msbar:Nf2}.

Therefore, in order to specify the hopping parameters of the heavy flavour,
$\kaph$, which according to \eq{M_mqtil} for given $L_1/a=2L_0/a$ and 
$\beta=6/g_0^2$ correspond to a series of dimensionless RGI quark mass values 
$z\equiv L_1M$ in the b-quark region, it remains to accurately determine the 
improvement coefficient $\bm$ and the renormalization constant $Z$. 
We discuss this computation for the relevant weak coupling range
(cf.~\tab{tab:bXzres}) of $\Or(a)$ improved two-flavour lattice QCD in the
next section.
\subsection{Preparing for the finite-volume computations}
\subsubsection{Matching to QCD}
As mentioned in the foregoing subsection, on the QCD side this step consists 
in calculating the quark mass dependence of the quantities $\Phi_1$, 
$\Phi_2$ and $\Phi_3$ in the volume $L_1^4$.
For the exact definitions of these effective heavy-light meson energies in
terms of SF correlators we refer to~\Ref{HQET:mb1m}.
In addition to $L_1^3\times T$, $T=L_1$, lattices with $T=L_1/2$ will also 
be needed (cf.~Appendix~C of \cite{HQET:mb1m}).

The aforesaid fine-tuning of $\beta$ for $L_1/(2a)\le 16$ to satisfy the 
condition $\gbsq(L_1/2)=2.989$, \eq{cond_L1}, with a precision 
$\Delta\gbsq\lesssim 0.04$ requires up to $(L_1/a)^4=(2L_0/a)^4=40^4$ 
lattices with $N_{\rm f}=2$ at sea quark parameters close to those quoted 
in \tab{tab:bXzres} in order to reach the continuum limit.
\subsubsection{%
Parameters for HQET simulations in $L_1^3\times T$ with $T=L_1,L_1/2$}
For the determination of the step scaling functions 
$L_1\rightarrow L_2=2L_1$ belonging to the $\Phi_k$'s counterparts in HQET, 
we must fix the simulation parameters for resolutions $6\le L_1/a\le 16$.
The corresponding constraint on the renormalized coupling at $L_1$ is
$\gbsq(L_1)=\sigma(2.989)=4.484(48)$~\cite{alpha:Nf2_2}.

As a starting point for the tuning of $\gbsq(L_1)$ at each $L_1/a$, we
introduce another low-energy scale, $L^{\ast}$, defined via 
$\gbsq(L^{\ast})=5.5$ and obeying \cite{lat07:rainer}
\be
\ln\left(L^{\ast}/a\right)=
2.3338+1.4025\left(\beta-5.5\right)\,,\quad
\beta\in[\,5.3\,,\,5.8\,]\,, \quad L^{\ast}/a\in[\,7.8\,,\,16.1\,]\,,
\label{param_Lstar}
\ee
which allows to estimate the ratio $r_1=L_1/L^{\ast}\approx 0.8$ in the 
continuum limit.
Trial $\beta$--estimates for the range of $L_1/a$ in question are then 
obtained from the parameterization (\ref{param_Lstar}) and improved by 
further simulations, aiming at a precision of $\Delta\gbsq\lesssim 0.1$.
This will be finished soon.

Small mismatches of the simulation results w.r.t.~the target values, 
i.e.~$\gbsq(L_1)=4.484$ and $L_1\ml(L_1)=0$, may be corrected by the 
non-perturbative $\beta$--function and the mass derivative of the 
coupling~\cite{alpha:Nf2_2,lat07:rainer}.
\subsubsection{%
Parameters for HQET simulations in $L_2^3\times T$ with $T=L_2,L_2/2$}
To prepare for the power-divergent subtractions in the volume of extent 
$L_2=2L_1\approx 1\,\Fm$ within the effective theory that eventually
provide the link to HQET observables in the physically large volume 
(of extent $L_{\infty}$), the two-flavour theory will have to be simulated 
at typical resolutions of about $L_2/a=8,12,16$ and lattice spacings 
corresponding to $5.3\lesssim\beta\lesssim 5.9$.

For fixing the necessary simulation parameters by means of the condition of
fixed coupling $\gbsq(L_2)$, one can rely again on the scale $L^{\ast}$ and
its ratio to $L_2$, $r_2=L_2/L^{\ast}|_{\,\rm continuum}\approx 1.6$, and infer
the wanted pairs $(L_2/a,\beta)$ from \eq{param_Lstar}.
\section{%
Computation of the missing improvement and renormalization factors}
\label{Sec_bXz}
We now present our non-perturbative determination of the improvement 
coefficient $\bm$ and the renormalization constant $Z$ in the $\beta$--range 
relevant for the matching of HQET to QCD in small volume, such that the RGI 
heavy quark mass can be set to desired values $z=L_1M$.

Our generation of unquenched gauge configurations with SF boundary 
conditions for $\nf=2$ $\Or(a)$ improved massless Wilson quarks employs the 
hybrid Monte Carlo (HMC) algorithm~\cite{hmc:orig} in its variant used in 
the study of autocorrelation times in \Ref{Nf2SF:autocorr}.
It comprises multi-time-scale integration schemes~\cite{hmc:mtsi1,hmc:mtsi2} 
with mass preconditioning~\cite{hmc:hasenb1,hmc:hasenb2,Nf2SF:algo} on top
of even-odd preconditioning.
%
\subsection{%
Non-degenerate current quark masses and estimators for 
$\ba-\bP$, $\bm$ and $Z$}
We proceed following the idea of imposing improvement conditions at constant 
physics, which was first advocated in~\cite{impr:babp} and already applied 
to the present situation but for $\nf=0$ in \cite{HQET:pap2}.

Since the definition (\ref{cond_L1}) of $L_1$ via the renormalized coupling
$\gbsq(L_1/2)=\gbsq(L_0)$, respectively the bare parameters in 
columns~2~--~4  of \tab{tab:bXzres} complying with it, have such a constant 
physics condition built in from the start, we can directly work at those 
pairs of $(L_0/a,\beta)$.
With this as our choice of improvement condition, supplemented by the 
SF-specific settings of zero boundary fields, $\theta=0.5$ and --- just for 
the purpose of this section --- $T/L_0=3/2$, the improvement coefficients 
$\ba-\bP$ and $\bm$ and the renormalization constant $Z$ become smooth 
functions of $g_0^2$ in the region where they are needed.\footnote{
Although the difference of coefficients $\ba-\bP$ is actually not needed for
fixing the RGI mass through \eq{M_mqtil}, we include it in the present 
discussion.
}

Taking over any unexplained notations and details 
from \Refs{HQET:pap2,impr:babp} (and references therein), $\ba-\bP$, $\bm$ 
and $Z$ can be determined by studying QCD with non-degenerate valence 
quarks. 
Treating the latter in the quenched approximation, the structure of the
$\Or(a)$ improved theory in conjunction with a massless renormalization 
scheme retains the relative simplicity of the $\nf=0$ case elaborated in 
\Ref{impr:babp}.
For instance, the improvement of the off-diagonal bilinear fields 
$X^{\pm}=X^1\pm i\,X^2$, $X=A_{\mu},P$, emerging as a consequence of the 
broken isospin symmetry in flavour space, is the same as in the degenerate 
case, except that the $b$--coefficients now multiply the average 
$\half(a\mqi+a\mqj)$ of the subtracted bare quark masses, $\mqi=\mzi-\mc$, 
which themselves are separately improved for each quark flavour:
\be
\mqitil\,=\,\mqi\left(1+\bm\,a\mqi\right)\,.
\ee 
(Here and below the indices $i,j$ label the different quark flavours.)
Identifying the valence flavours in the isospin doublet with a light 
(strange) and a heavy (bottom) quark, the corresponding PCAC relation reads
\be
\Pmu A_{\mu}^{\pm}(x)=(m_i+m_j)P^{\pm}(x)\,,
\ee
and the renormalization constants $\za$ and $\zp$ that come into play upon 
renormalization are just those known in the theory with two mass-degenerate 
quarks.

Accordingly, the SF correlation functions involving the axial current and 
the pseudoscalar density generalize to 
$\fa^{ij}(x_0)=-\frac{1}{2}\mvl{A_0^+(x)O^-}$ and 
$\fp^{ij}(x_0)=-\frac{1}{2}\mvl{P^+(x)O^-}$, with pseudoscalar boundary
sources decomposed as $O^{\pm}=O^{1}\pm i\,O^{2}$ where
$O^{a}=a^6\sum_{\mby,\mbz}\zeb(\mby)\gfv\,\gen\,\zeta(\mbz)$.
Then the improved bare PCAC (current) quark masses\footnote{
This expression for the PCAC masses is only $\Or(a)$ improved up to a factor
$1+\half(\ba-\bP)(a\mqi+a\mqj)$ for quark mass dependent cutoff effects. 
}
as functions of the timeslice location $x_0$ are given by
\be
m_{ij}(x_0;L_0/a,T/L_0,\theta)=
\frac{\Sz\fa^{ij}(x_0)+a\ca\drvstar{0}\drv{0}\fp^{ij}(x_0)}
{2\,\fp^{ij}(x_0)}\,,
\label{mpcac_x0}
\ee
where only here we explicitly indicate their additional dependence on 
$L_0/a$, $T/L_0$ and $\theta$.
In the degenerate case, $i=j$, the correlators assume the standard form, 
and $m_{ij}$ just reduces to the current quark mass of a single quark flavour 
that is prepared by a corresponding choice of equal values for the 
associated hopping parameters, $\kappa_i=\kappa_j$.
Also the precise definition of the lattice derivatives in \eq{mpcac_x0}
matters.
As it is written there, $\Sz=\half(\drv{0}+\drvstar{0})$ denotes the average
of the ordinary forward and backward derivatives, but as 
in \Refs{HQET:pap2,impr:babp} we have used the improved derivatives 
\be
\Sz\rightarrow
\Sz\left(1-{\T \frac{1}{6}}\,a^2\drvstar{0}\drv{0}\right)\,,\quad
\drvstar{0}\drv{0}\rightarrow
\drvstar{0}\drv{0}\left(1-{\T \frac{1}{12}}\,a^2\drvstar{0}\drv{0}\right)
\label{deriv}
\ee
as well, which (when acting on smooth functions) have $\Or(g_0^2a^2,a^4)$ 
errors only.

For their numerical calculation, the coefficients $\ba-\bP$, $\bm$ and the 
finite factor $Z=\zm\zp/\za$ (see~\eq{Zm_mqh}) are isolated by virtue of 
the identity
\be
m_{ij}=
Z\,\Big[\,
\half\left(\mqi+\mqj\right)+\half\,\bm\left(a\mqi^2+a\mqj^2\right)
-\,\quart\left(\ba-\bP\right)a\left(\mqi+\mqj\right)^2
\,\Big]\,+\,\Or\left(a^2\right)\,.
\label{mpcac_mq}
\ee
It is obtained by equating the expression for the $\Or(a)$ improved 
renormalized quark mass in terms of the bare PCAC mass with the alternative 
expression in terms of the subtracted bare quark mass.
Forming ratios of suitable combinations of degenerate and non-degenerate 
current quark masses in the representation (\ref{mpcac_mq}) then enables to
derive direct estimators for $\ba-\bP$, $\bm$ and $Z$ \cite{impr:babp}:
\bea
R_{\rm AP}
& = &
\frac{2\,(2m_{12}-m_{11}-m_{22})}
{(m_{11}-m_{22})(am_{{\rm q},1}-am_{{\rm q},2})}
\,\,=\,\,\ba-\bP\,+\,\Or\left(am_{{\rm q},1}+am_{{\rm q},2}\right)\,,
\label{estim_bap}\\
R_{\rm m}
& = &
\frac{4\,(m_{12}-m_{33})}
{(m_{11}-m_{22})(am_{{\rm q},1}-am_{{\rm q},2})}
\,\,=\,\,\bm\,+\,\Or\left(am_{{\rm q},1}+am_{{\rm q},2}\right)\,,
\label{estim_bm}
\eea
with $m_{0,3}=\half(m_{0,1}+m_{0,2})$, neglecting other quark mass independent 
lattice artifacts of $\Or(a)$.
For the renormalization constant $Z$ an analogous formula holds even up to 
$\Or(a^2)$ corrections,
\be
R_{Z}\,\,=\,\,
\frac{m_{11}-m_{22}}{m_{{\rm q},1}-m_{{\rm q},2}}
\,+\,(\ba-\bP-\bm)(am_{11}+am_{22})\,\,=\,\,
Z\,+\,\Or\left(a^2\right)\,,
\label{estim_Z}
\ee
if the correct value for $\ba-\bP-\bm=R_{\rm AP}-R_{\rm m}$ (only involving 
correlation functions with mass degenerate quarks) is inserted.
Note that generically the combination $Z=\zm\zp/\za$ is a function of the 
improved bare coupling, $\tilde{g}_0^2=g_0^2\,(1+b_{\rm g}\,a\mq)$.
Since, however, we only consider light sea quarks that are massless 
(i.e.~such that $\ml\approx 0$) and the valence quarks are anyway treated in 
the quenched approximation, this fact can be ignored here.

Still, to complete our definition of the line of constant physics, values 
for the bare PCAC masses of the valence quarks must be selected. 
As in \cite{HQET:pap2}, we consider two pairs,
\bea
\mbox{choice~1}:
& \quad &
L_0 m_{11}\approx 0 \,,\quad L_0 m_{22}\approx 0.5\,,
\label{choice_m1}\\
\mbox{choice~2}:
& \quad &
L_0 m_{11}\approx 0 \,,\quad L_0 m_{22}\approx 2.4\,.
\label{choice_m2}
\eea
The first choice on $L_0 m_{22}$ is motivated by the quenched 
investigation~\cite{impr:babp}, where it was argued to be advantageous 
w.r.t.~the size of $\Or(a)$ ambiguities encountered, while with the second  
choice one is closer to the typical b-quark region itself.
Satisfying these conditions on $L_0 m_{22}$ for all $(L_0/a,\beta)$ 
in~\tab{tab:bXzres} demands to properly adjust the hopping parameter, called
$\kaph$ above, that is responsible for the mass value of the heavy valence 
quark flavour.
This in turn amounts to prior evaluations of the relevant correlation 
functions on the dynamical gauge background for some trial guesses of 
$\kaph$, in order to estimate the heavy flavour's PCAC mass 
through \eq{mpcac_x0} and to tune it to the values dictated by 
\eqs{choice_m1} and (\ref{choice_m2}) up to a few percent.\footnote{
Similar to the situation in \Refs{HQET:pap2,impr:babp}, this is to 
sufficient precision equivalent to keeping fixed the corresponding 
renormalized masses $L_0\za m/\zp$, as for the considered couplings the 
entering renormalization constant barely varies.
}
The resulting hopping parameters are given in the fifth column 
of~\tab{tab:bXzres}.
\subsection{Results}
%
%
\begin{table}[t]
\centering
\small
\renewcommand{\arraystretch}{1.25}
\begin{tabular}{cCCCCCCCC} 
\toprule
  set & L_0/a & \beta & \kapl & \kaph &  L_0 m_{22} & \ba\!-\!\bp 
& \bm & Z \\ 
\midrule
  1 & 10 & 6.1906  & 0.136016 & 0.134318 &  0.4929(6)  & -0.0006(9) 
& -0.6643(9) & 1.1046(1) \\
    & 12 & 6.3158  & 0.135793 & 0.134378 &  0.4952(9)  & -0.003(2) 
& -0.668(2)  & 1.1050(2) \\
    & 16 & 6.5113  & 0.135441 & 0.134387 &  0.492(1)   & -0.006(2) 
& -0.667(3)  & 1.1044(2) \\
    & 20 & 6.6380  & 0.135163 & 0.134356 &  0.5005(9)  & -0.005(3) 
& -0.669(3)  & 1.1038(2) \\
\midrule
  2 & 10 & 6.1906  & 0.136016 & 0.127622 &  2.2909(5)  & +0.0727(4)  
& -0.5655(3) & 1.0954(1) \\
    & 12 & 6.3158  & 0.135793 & 0.128755 &  2.3475(7)  & +0.0513(5)  
& -0.5785(5) & 1.0974(1) \\
    & 16 & 6.5113  & 0.135441 & 0.130146 &  2.407(1)   & +0.0297(7)  
& -0.5964(8) & 1.0995(1) \\
    & 20 & 6.6380  & 0.135163 & 0.130965 &  2.4433(8)  & +0.0215(6)  
& -0.6076(8) & 1.1002(1) \\
\bottomrule
\end{tabular}
\caption{%
Lattice parameters and numerical results on the improvement coefficients 
$\ba-\bP$ and $\bm$ and on the renormalization constant $Z$.
The parameters $(L_0/a,\beta,\kapl)$ referring to the light (sea) quark 
sector have fixed SF coupling, $\gbsq(L_0)=2.9(1)$, and vanishing quark mass 
such as to meet the constant physics condition 
of~\protect\Sect{Sec_survey_L1}.
Our results for $\ba-\bP$, $\bm$ and $Z$ are based on statistics varying 
from $\Or(300)$ measurements ($L_0/a=20$) to $\Or(2000)$ measurements 
($L_0/a=10$). 
The upper set refers to ``choice~1'', \protect\eq{choice_m1}, where the 
heavy quark mass is kept at $L_0 m_{22}\approx 0.5$, while the lower set 
belongs to ``choice~2'' with $L_0 m_{22}\approx 2.4$, \protect\eq{choice_m2}.
The condition $L_0 m_{11}\approx 0$ is fulfilled up to negligible deviations 
of about $0.015$ at most.
}\label{tab:bXzres}
\end{table}
%
%
\begin{figure}[t]
\centering
\leavevmode
\epsfig{file=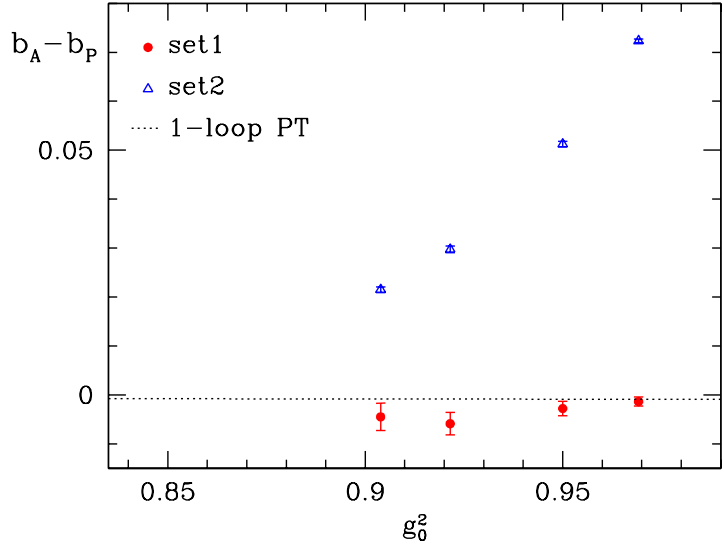,width=0.625\textwidth}
\caption{%
Two sets of non-perturbative results for $\ba-\bP$ in the considered region 
of bare couplings, referring to our two choices of quark masses, together 
with the prediction from one-loop perturbation theory.
}\label{fig:bap}
\end{figure}
The technical aspects of the analysis to compute the estimators 
(\ref{estim_bap}) -- (\ref{estim_Z}) from the numerical data on the 
heavy-light SF correlation functions by means of the PCAC masses $m_{ij}$ 
for the various (degenerate and non-degenerate) valence quark mass 
combinations are the same as in \Refs{HQET:pap2,impr:babp}.
The correlators have been evaluated on our dynamical gauge field 
configurations, which were generated on $L_0^3\times T$ lattices with 
$T=3L_0/2$ and massless sea quarks (thus complying with the above 
requirement $\kapl=\kapc$ resp. $L_0 m_{11}\approx 0$ for the light valence 
quark flavour) and which were separated by 5~--~10 HMC trajectories of 
length one.
As for the $m_{ij}$ themselves, they have been calculated from the local 
masses, \eq{mpcac_x0}, using improved derivatives (\ref{deriv}) throughout
and averaging over the central timeslices $L_0/(2a),\ldots,(T-L_0/2)/a$ to 
increase statistics.
Being secondary quantities in particular, the statistical errors of the
masses and of the $R_{\rm X}$, ${\rm X}={\rm AP},{\rm m},Z$, obtained from 
them were estimated by the $\Gamma$--method \cite{MCerr:ulli}, which 
directly analyzes autocorrelation functions.

\begin{figure}[t]
\centering
\leavevmode
\epsfig{file=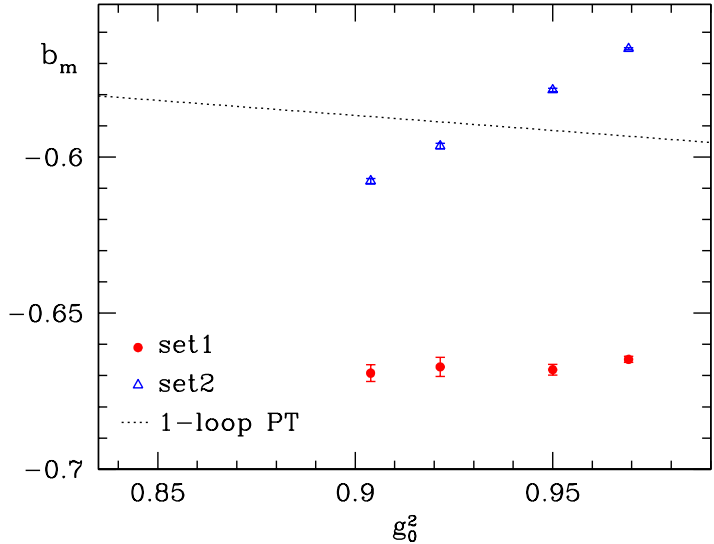,width=0.625\textwidth}
\caption{%
The same as in \protect\fig{fig:bap} but for the improvement coefficient 
$\bm$.
}\label{fig:bm}
\end{figure}

\begin{figure}[t]
\centering
\leavevmode
\epsfig{file=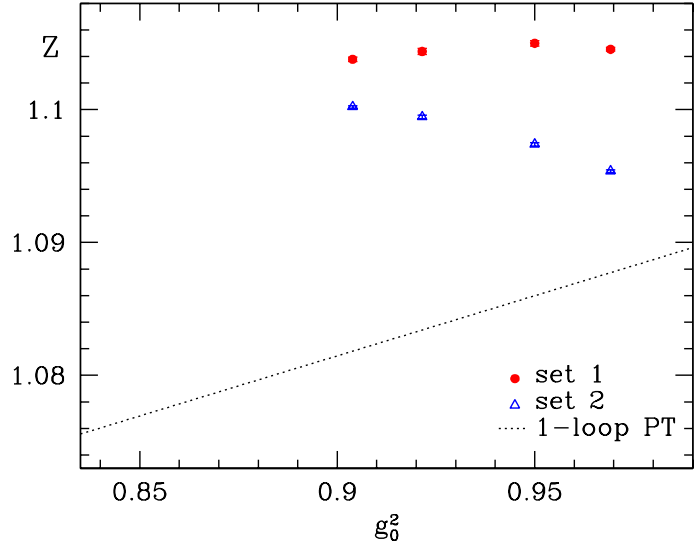,width=0.625\textwidth}
\caption{%
The same as in \protect\fig{fig:bap} but for the renormalization constant 
$Z$.
}\label{fig:Z}
\end{figure}

Our non-perturbative results on $\ba-\bP$, $\bm$ and $Z$ are also listed 
in~\tab{tab:bXzres}.
As a consequence of the underlying constant physics 
condition (\ref{cond_L1}), the estimates $R_{\rm X}$, 
${\rm X}={\rm AP},{\rm m},Z$, become smooth functions of the bare coupling, 
$g_0^2=6/\beta$.
This is well reflected in \figs{fig:bap} -- \ref{fig:Z}, where our results 
are shown in comparison with the one-loop perturbative 
predictions~\cite{impr:babp,impr:pap5}.

%
%
%
%
The overall $g_0^2$--dependence of our results is qualitatively similar to 
the quenched study \cite{HQET:pap2} and even comparable on the quantitative
level.
Whereas $R_{\rm AP}$ is compatible with a nearly vanishing $\ba-\bP$, as 
predicted by leading-order perturbation theory, for ``choice~1'' of quark 
masses and appears to approach this line quite rapidly as 
$g_0^2\rightarrow 0$ for ``choice~2'', one observes for both choices 
significant deviations of the sets of estimates for $\bm$ and $Z$ from the 
leading perturbative behaviour in the weak coupling region considered.
Since one expects the perturbative curves eventually to be approached in the 
limit $g_0^2\rightarrow 0$ also in case of $\bm$ and $Z$, the curvature seen 
in our numbers hints at a more complicated structure of (unknown) 
higher-order terms.
Hence, we have to conclude that if an improvement condition were used in a 
region of stronger couplings, which would no doubt lead to a rather 
different set of data points, simple one-loop perturbation theory would not 
be an adequate guide for the continuation of $\bm$ and $Z$ to weak 
couplings.
On the contrary, this would induce a source of uncertainty in results 
deriving from them that is difficult to control and, therefore, highlights 
the importance of employing improvement conditions in the $\beta$--range 
relevant to the actual application.

%
%
\begin{figure}[t]
\centering
\leavevmode
\epsfig{file=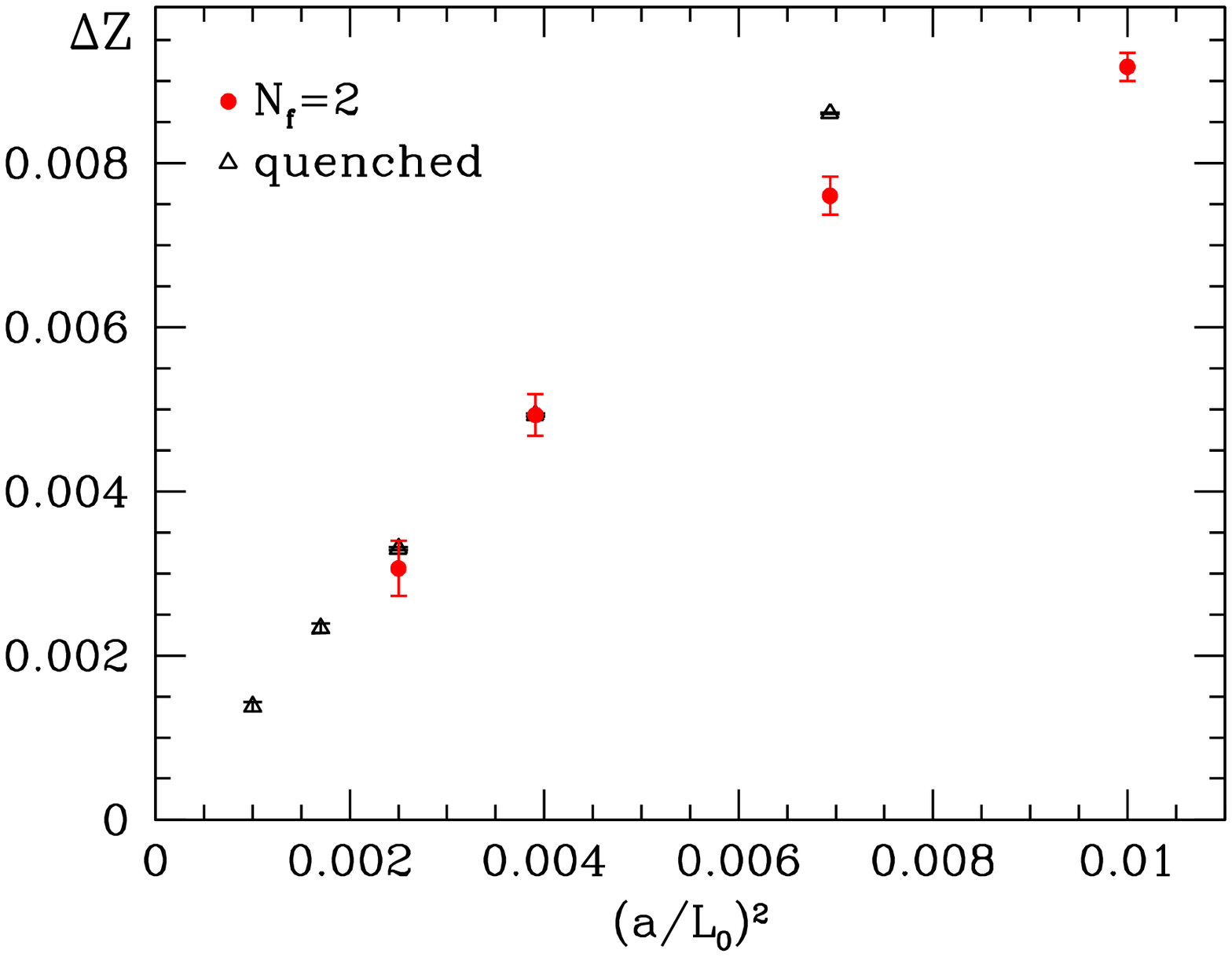,width=0.49\textwidth}
\epsfig{file=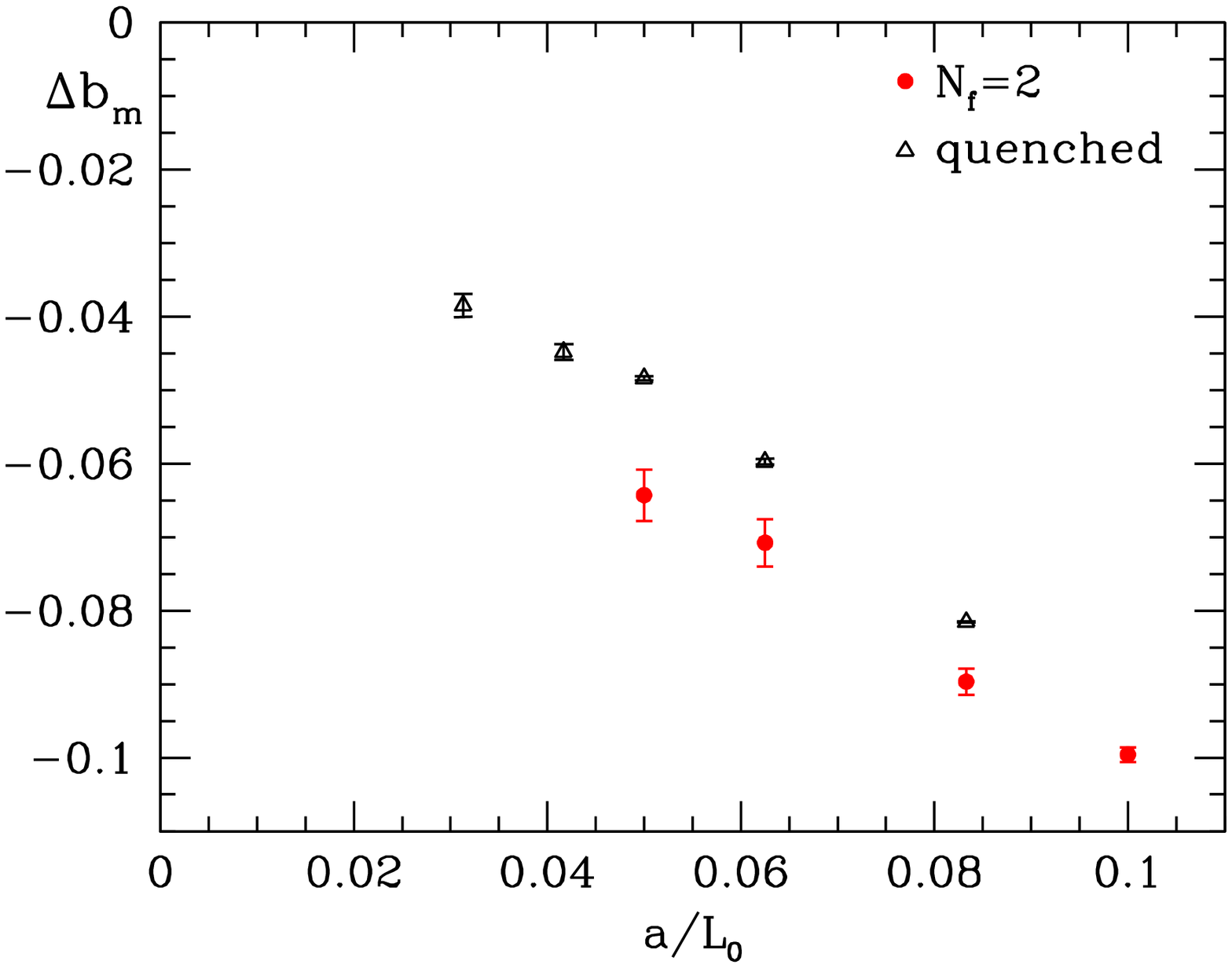,width=0.49\textwidth}
\caption{%
Left: Difference of the two sets of results in \protect\tab{tab:bXzres} on 
the renormalization constant $Z$ versus $(a/L_0)^2$.
Right: The same for the improvement coefficient $\bm$ where, however,
the ambiguity inherent in any improvement condition imposed is of $\Or(a)$.
The open black triangles display the corresponding quenched results 
from \protect\cite{HQET:pap2} for comparison.
}\label{fig:delZbm}
\end{figure}
Of course, any other estimate $R_{\rm X}$ (i.e.~stemming from a different 
choice of renormalization/improvement condition) may yield a different 
functional dependence upon $g_0^2$, but its differences are again smooth 
functions that vanish in the continuum limit with a rate proportional to 
$a/L_0$ (for improvement coefficients) or $(a/L_0)^2$ (for renormalization 
constants).
These intrinsic $\Or(a^n)$ ambiguities ($n=1,2$) imply that rather than a 
numerical value at some given $\beta$, the important information lies in the 
correct \emph{$g_0^2$--dependence} of the estimators $R_{\rm X}$, 
${\rm X}={\rm AP},{\rm m},Z$, obtained at \emph{constant physics}.
To demonstrate this, we also investigated a few alternative improvement 
conditions, which are either provided by defining the estimators $R_{\rm X}$ 
with standard instead of improved derivatives or by the two quark mass
choices, \eqs{choice_m1} and (\ref{choice_m2}), themselves.
As an example we plot in the left panel of \fig{fig:delZbm} the difference
$\Delta Z(g_0^2)=Z(g_0^2)|_{\,\rm choice\,1}-Z(g_0^2)|_{\,\rm choice\,2}$ 
versus $(a/L_0)^2$, which clearly exhibits a linear approach towards zero.
Other cases behave similarly, e.g.~the $\Or(a)$ ambiguities for 
$\Delta\bm(g_0^2)=
\bm(g_0^2)|_{\,\rm choice\,1}-\bm(g_0^2)|_{\,\rm choice\,2}$ in the right 
panel of \fig{fig:delZbm} are found to be quite small, and their magnitude 
rapidly decreases as $a/L_0\rightarrow 0$.
\section{Outlook}
\label{Sec_outl}
Apart from the elements sketched at the end of \Sect{Sec_survey}, which 
partly are already in progress, the computation of the b-quark mass at the 
$\minv$--order of HQET along our strategy illustrated in \fig{fig:strat} 
still requires $\nf=2$ simulations in $L_2\approx 1\,\Fm$ as well as in 
physically large volumes of about $L_{\infty}\gtrsim 2\,\Fm$.
Particularly for the latter we plan to switch to QCD with periodic boundary 
conditions and to use the technique of low-mode 
deflation~\cite{deflat:luescher} in connection with all-to-all quark 
propagators~\cite{ata:dublin} for the numerical evaluation of correlation 
functions.

As further interesting directions for future work let us mention the
non-perturbative tests of the HQET expansion in the spirit 
of~\Ref{HQET:pap3} and the extension of our determination of improvement 
coefficients and $Z$--factors to the parameter range relevant for (large
volume) charm physics.
%

\vspace{0.5cm}
\noindent {\bf Acknowledgements.}
We thank NIC for allocating computer time on the APE computers to this 
project and the APE group at Zeuthen for its support.
We further acknowledge partial support~by the Deutsche 
Forschungsgemeinschaft (DFG) in the SFB/TR 09-03,
``Computational Particle Physics'', and by the European Community through 
EU Contract No.~MRTN-CT-2006-035482, ``FLAVIAnet''.
P.F. and J.H. also acknowledge financial support by the DFG under grant
HE~4517/2-1.

%
\bibliography{lattice_ALPHA}
\bibliographystyle{JHEP-2}
\end{document}